\documentclass[a4paper,11pt]{article}
\usepackage{pos}

\title{A tale of two SUSYs}

\author*[a]{Stam Nicolis}

\affiliation[a]{Institut Denis Poisson (CNRS UMR 7013),\\
Université de Tours, Parc Grandmont, 37200 Tours, France}

\emailAdd{stamatios.nicolis@univ-tours.fr}

\abstract{There are two approaches towards supersymmetry: The ``conventional approach'', in which the fields appear in the classical action and the ``stochastic approach'', in which they emerge upon introducing in the action the contribution of a certain determinant. The second approach relies, in particular, on the so-called Nicolai map. The relation between the two approaches hasn't been clarified and the subject of this contribution is to spell it out. In particular the interpretation proposed by Parisi and Sourlas has remained quite incomplete. The subject of this contribution is to spell out in what way the two approaches complement each other and how the second can provide insights for the first.}

\FullConference{Proceedings of the Corfu Summer Institute 2024 "School and Workshops on Elementary Particle Physics and Gravity" (CORFU2024)\
12 - 26 May, and 25 August - 27 September, 2024\\
Corfu, Greece\\}

\begin{document}
\maketitle

\section{Introduction}\label{intro}
There are two approaches towards supersymmetry: The first, ``conventional'', approach treats supersymmetry like any other continuous symmetry and focuses on the construction of supermultiplets and culminates with the definition of superfields, that act in superspace. 
While this approach has been pursued since the discovery that supersymmetry could be defined as a symmetry of relativistic particles and fields and much progress has been made in understanding the structures involved, much, still,  remains to be understood, both from the mathematical point of view as well as from that of applications; similar to what is the case for any quantum field theory, only, in the case of supersymmetric quantum field theories the constraints are much more challenging to take into account.

Another approach towards the description of supersymmetric theories was pioneered by Nicolai~\cite{Nicolai:1980jc,Nicolai:1980js}. 
The starting point was the realization that fermionic degrees of freedom posed particular problems when trying to go beyond perturbation theory, namely using a lattice regularization~\cite{Dondi:1976tx}; therefore it might be useful to integrate them out and deal with the determinant they represent in a different way, namely by means of what has been since become known as the ``Nicolai map'', an expression of only the bosonic superpartners of the fermions. How to construct this map is a non-trivial task; the attempts that were undertaken in the 1980s~\cite{deAlfaro:1982ex,Flume:1983sx,deAlfaro:1984hb,deAlfaro:1984gw,Dietz:1984hf,DeAlfaro:1986uv,Dietz:1985hga,Lechtenfeld:1986gd}, indeed, seemed to hit non-trivial conceptual and technical obstacles. In particular, the map was constructed order by order in perturbation theory, but the corresponding ``Feynman rules'' are, still, missing. 

Recently, this effort has been revived~\cite{Ananth:2020lup,Ananth:2020gkt,Lechtenfeld:2021yjb,Lechtenfeld:2021uvs,Lechtenfeld:2021zgd,Lechtenfeld:2022qpa,Lechtenfeld:2022fxs,Lechtenfeld:2022qed,Lechtenfeld:2022lvb,Casarin:2023xic,Lechtenfeld:2024uhi,Lechtenfeld:2024ecf}; a  review  may be found in ref.~\cite{Lechtenfeld:2023gcq}. 

What is missing in this effort is whether there is some physical insight behind what the Nicolai map represents, or is it, simply, a technical framework for organizing certain calculations in supersymmetric theories. The physical insight was  hinted to by Nicolai's first papers,  advanced further  in the papers by de Alfaro {\em et al.} (where the term ``stochastic'' is used); but, not, really, spelled out. Furthermore, the rules for constructing the Nicolai map--even perturbatively--are, still, very far from the simplicity  of the ``Feynman rules'' as these have become to be understood in quantum field theory, so there's still much to understand about the Nicolai map, already at this level.

In 1982 Parisi and Sourlas~\cite{parisi_sourlas}, finally, spelled out what the physical insight was--and showed that supersymmetry has a much broader scope than hitherto acknowledged. They argued, in effect, that the Nicolai map represented the correspondence between the physical degrees of freedom of any physical system and the ``noise fields'', describing the bath of fluctuations, with which the physical degrees of freedom were in equilibrium. In particular, they showed that, in two dimensional, Euclidian, spacetime, 
if the physical degrees of freedom were two scalars, then the noise fields, defined by the Nicolai map, lead to the appearance of the 
determinant identified  by Nicolai, upon starting from a supersymmetric theory, in such a way that, when introduced into the action of the scalars through anticommuting fields, the full  action could be identified with that of  the $\mathcal{N}=2$ Wess-Zumino model--even though the starting point was a theory without any apparent supersymmetry whatsoever! They further argued that this implied that the fermionic superpartners provided a resolution of the degrees of freedom that constitute the bath of fluctuations for the scalars--and that this holds, in fact, whatever the nature of the bath of fluctuations.  It is this identification that provides the justification for the qualifier ``stochastic'', used for characterizing this approach to supersymmetry.

The difference between the two approaches to supersymmetry can, therefore, be expressed in the following way: 
The conventional approach to supersymmetry starts from the classical action, that incorporates all the fields and their combinations, allowed by the symmetries (namely invariance under Lorentz transformations and/or rotations and translations of the worldvolume coordinates and invariance under supersymmetry transformations; and the induced transformations on the fields themselves); the stochastic approach leverages the property that certain fields may ``emerge''--which is, in fact, what happens with fermions, that allow  taking into account the effects of non-local degrees of freedom, namely determinants of local operators, as local contributions in the action. The question then arises, how to take into account the contribution of this determinant. In perturbation theory, of course, the Feynman rules for fermions are unambiguous; beyond perturbation theory, e.g. using a lattice regularization, things are more complicated, since the determinant need not be positive definite and is, certainly, non-local. These are the main obstacles for making full and/or direct use of the Nicolai map in lattice simulations~\cite{Schaich:2018mmv,Joseph:2023vja,Schaich:2024bmg}. 

The plan of the paper is thus, the following: In section~\ref{tsSUSY} we recall the ``conventional'' approach towards supersymmetry and the obstacles it encounters; in section~\ref{Nicolaimap} we review the idea behind the Nicolai map, especially in view of its recent revival and what are-some-of the issues it faces for further progress; in section~\ref{stochSUSY} we review the insight of Parisi and Sourlas and the issues it faces. Our conclusions are summarized in section~\ref{concl}.

\section{Target space SUSY}\label{tsSUSY}
Now it should be stressed that the supersymmetry that we have been discussing up to now hides some assumptions, that it is useful to spell out. 

 It should be recalled that fields are maps from a ``worldvolume'' to a ``target space''. In applications to particle physics, the worldvolume is flat spacetime, with the Minkowski metric; for many purposes it is useful to work, also, in Euclidian signature. 
 The target space is the space the fields take values in. But it, also, has a geometric structure, related to that of the worldvolume. 
 In the ``conventional'' approach to supersymmetry, the fields transform as scalars, vectors and spinors under Lorentz transformations of the worldvolume coordinates (or in Euclidian signature, depending on the context). So they carry ``target space'' indices and the supersymmetry they realize is ``target space supersymmetry''. 
 
 These fields, in turn, can be repackaged as ``superfields'', maps from the superspace of the worldvolume, augmented by anticommuting coordinates, to the superspace of fields. 
 
 The problem with this approach is that, for extended supersymmetry, how to take into account the constraints is not known to full generality. Now it might be thought that the ``simplest'' realization of supersymmetry, $\mathcal{N}=1$ supersymmetry, ought to be both necessary and sufficient for applications, since its superspace formalism is known~\cite{Gates:1983nr} and it can describe chiral fermions. However, as is well known, the simplicity of $\mathcal{N}=1$ supersymmetry is illusory and how its properties, that are desirable for applications to particle physics, can emerge remains to be understood. Furthermore, while how supersymmetry can be realized--at least in perturbation theory about free fields--is understood, how it is broken, is, still, not, fully, understood. Nor is its place, in the context of particle physics. Indeed, insight into model-building has come from the study of extended supersymmetry, that provides a much better framework for carrying out reliable calculations. Providing constraints for experimental searches, however, is, still, very much, work in progress. The limits obtained by the collaborations at the LHC~\cite{ATLAS:2024fub,Rathjens:2024hlj} constrain  the attempt to parametrize ways of breaking supersymmetry explicitly, more than anything else. So they eliminate many ways of breaking it by hand; while this is, of course, useful, it doesn't shed light on how to understand the possible ways it can be broken (that it must be broken is, of course, obvious, since we do not observe superpartners of the known particles with the same mass). 
 
 So it might appear that this approach towards understanding supersymmetry seems to reach an obstacle towards further understanding its role. While it is, of course, possible to pursue calculations, especially harnessing contemporary computing resources, the question remains what insights are brought to bear. 
 
 It is at this point that the insight of Parisi and Sourlas--and of Nicolai--becomes relevant. It allows us to understand supersymmetry's role in a different way. This is the subject of the following section.
 
\section{Intermezzo: The Nicolai map}\label{Nicolaimap}
The Nicolai map, according to the recent review~\cite{Lechtenfeld:2023gcq}, maps an interacting, non-local, theory of commuting fields, that is, however, equivalent to a supersymmetric theory (the fermionic superpartners have been integrated out), to a theory of free fields. 

This statement, however, is incomplete. The free fields are not just any free fields; they have ultra--local 2-point function. This point has been consistently left out of all considerations of the Nicolai map. This is, indeed, the observation made by Parisi and Soiurlas
 in their paper in 1982--though Parisi and Sourlas claimed that what they were discussing didn't have anything to do with the Nicolai map. The ``noise fields'' they are discussing are, precisely, the free fields of Nicolai and Lechtenfeld. 
 
 Parisi and Sourlas proposed to identify the noise fields by invoking the Langevin equation, because they were working within the framework of stochastic quantization, wherein the fluctuations are quantum fluctuations. However there is s subtle point here: The Langevin equation,
 \begin{equation}
 \label{Langevin}
 \eta_I=\frac{\partial\phi_I}{\partial\tau}+\frac{\partial S}{\partial\phi_I}
 \end{equation}
for field theories (cf.~\cite{ZinnJustin:2002ru}), describes the approach towards equilibrium of a system, defined by the Euclidian action $S[\phi_I]$; as noted by Parisi and Sourlas, this leads to a partition function for the fields $\phi_I$ that contains more than two derivatives, which is, also,  problematic. 

What they really meant (and their calculations are done within this framework) is that what they call-by abuse of language-the Langevin equation does not describe the {\em approach towards} equilibrium, but the mapping between the ``noise fields'', $\eta_I$ and the ``dynamical fields'', $\phi_I$ {\em at} equilibrium\footnote{I'm grateful to A. Schwimmer for stressing this distinction.}. So $\tau$ in the above equation is not the time towards equilibrium, but the 
Euclidian time, for a (non-relativistic) particle and $S$ is not the full Euclidian action, but the superpotential for the (non-relativistic) particle. 

So, for the non-relativistic particle, with $\eta_I=\eta_I(\tau)$ and $\phi_I=\phi_I(\tau)$ and setting $S=W(\phi_I)$ where $I=1,2,\ldots,d$ labels the target space of the particle, if we assume that the $\{\eta_I(\tau)\}$ are free fields with correlation functions
\begin{equation}
\label{etamoms}
\begin{array}{l}
\displaystyle
\langle\eta_I(\tau)\rangle = 0\\
\displaystyle
\langle\eta_I(\tau)\eta_J(\tau')\rangle = 2\delta_{IJ}\delta(\tau-\tau')
\end{array}
\end{equation} 
and the higher order correlation functions are defined by Wick's theorem,
\begin{equation}
\label{wick}
\langle\eta_{I_1}(\tau_1)\eta_{I_2}(\tau_2)\cdots\eta_{I_{2n}}(\tau_{2n})\rangle = 
\sum_\pi\,\langle\eta_{I_{\pi(1)}}(\tau_{\pi(1)})eta_{I_{\pi(2)}}(\tau_{\pi(2)})\rangle\cdots
\langle\eta_{I_{\pi(2n-1)}}(\tau_{\pi(2n-1)})\eta_{I_{\pi(2n)}}(\tau_{\pi(2n)})\rangle
\end{equation}
this means that the partition function for the noise fields is given by the expression
\begin{equation}
\label{Znoise}
Z_\mathrm{noise}=\int\,[\mathscr{D}\eta_I]\,e^{-\int\,d\tau\,\frac{1}{2}\eta_I(\tau)\eta_J(\tau)\delta^{IJ}}
\end{equation}
Now, if we take eq.~(\ref{Langevin}) precisely as the Nicolai map, then we realize that it is an injunction to perform a change of variables in the partition function and obtain the partition function for the scalars $\phi_I(\tau):$
\begin{equation}
\label{Zphi}
Z_\mathrm{noise}=Z_\phi=\int\,[\mathscr{D}\phi_I]\,\left|\mathrm{det}\,\frac{\delta\eta_I(\tau)}{\delta\phi_J(\tau')}\right|\,
e^{-\int\,d\tau\,\frac{1}{2}\left(\frac{\partial\phi_I}{\partial\tau}+\frac{\partial W}{\partial\phi_I}\right)
\left(\frac{\partial\phi_J}{\partial\tau}+\frac{\partial W}{\partial\phi_J}\right)\delta^{IJ}}
\end{equation}
and we immediately identify the Jacobian as the determinant of a local operator, namely,
\begin{equation}
\label{Nicolaidet}
\frac{\delta\eta_I(\tau)}{\delta\phi_J(\tau')}=\delta(\tau-\tau')
\left(\frac{d}{d\tau}\delta^{IJ}+\frac{\partial^2 W}{\partial\phi_I(\tau)\partial_J(\tau)}\right)
\end{equation}
It is this property that makes it useful to introduce the operator in the Euclidian action using Grassmann variables, $\psi_I(\tau)$ and $\chi_J(\tau),$ with $I,J=1,2,\ldots,d:$
\begin{equation}
\label{Grassmann}
\mathrm{det}\left( \delta(\tau-\tau')
\left(\frac{d}{d\tau}\delta^{IJ}+\frac{\partial^2 W}{\partial\phi_I(\tau)\partial_J(\tau)}\right)\right)=
\int\,[\mathscr{D}\psi_I(\tau)][\mathscr{D}\chi(\tau)]\,e^{\int\,d\tau\,\psi_I(\tau)\left\{\frac{d}{d\tau}\delta^{IJ}+\frac{\partial^2 W}{\partial\phi_I(\tau)\partial_J(\tau)}\right\}\chi_J(\tau) }
\end{equation}
Even better, we remark that it's the absolute value of the determinant of this operator that appears inevitably in the partition function for the scalars, that describe the position of the particle. Therefore the partition function for the scalars (the components of the position of the non-relativistic particle) that is obtained, upon performing the change of variables of eq.~(\ref{Langevin}), is given by the expression
\begin{equation}
\label{ZSUSY}
Z_\phi=\int\,[\mathscr{D}\phi_I][\mathscr{D}\psi_I][\mathscr{D}\chi_I]\,e^{-S[\phi_I,\psi_I,\chi_I]}\,e^{-\mathrm{i}\theta_\mathrm{det}}
\end{equation}
where $\theta_\mathrm{det}$ is the phase (in general) of the determinant of the local operator of eq.~(\ref{Nicolaidet}). Therefore we can identify this expression with the Witten index~\cite{Witten:1982df,Witten:1982im}; which is consistent with the fact that we may take $Z_\mathrm{noise}=1$ by an appropriate choice of units and the change of variables realized by the Nicolai map~(\ref{Langevin}) can't change the value of the integral--in the absence of anomalies. 

Now the action, that includes the contribution of the Jacobian,
\begin{equation}
\label{Ssusy}
S[\phi_I,\psi_I,\chi_I]=\int\,d\tau\,\left[
\frac{1}{2}\left(\frac{\partial\phi_I}{\partial\tau}+\frac{\partial W}{\partial\phi_I}\right)
\left(\frac{\partial\phi_J}{\partial\tau}+\frac{\partial W}{\partial\phi_J}\right)\delta^{IJ}-\psi_I(\tau)\left\{\frac{d}{d\tau}\delta^{IJ}+\frac{\partial^2 W}{\partial\phi_I(\tau)\partial_J(\tau)}\right\}\chi_J(\tau)\right]
\end{equation}
(but not its phase) can be shown to be invariant, up to to total derivatives, under the transformations
\begin{equation}
\label{SUSYtransfs}
\begin{array}{lcl}
\displaystyle
\delta_{\zeta_1}\phi_I = -\zeta_1\chi_I & & \delta_{\zeta_2}\phi_I = \zeta_2\psi_I  \\
\displaystyle
\delta_{\zeta_1}\psi_I =  -\zeta_1\left(\dot{\phi}_I + \frac{\partial W}{\partial\phi_I}\right)& &  \delta_{\zeta_2}\psi_I =0\\
\displaystyle
\delta_{\zeta_1}\chi_I= 0 & & 
\displaystyle
\delta_{\zeta_2}\chi_I =\zeta_2\left(\dot{\phi}_I-\frac{\partial W}{\partial\phi_I}\right) \\
\end{array}
\end{equation}
where the $\zeta_{1,2}$ are Grassmann variables,
that close on the translations--hence the justifcation for calling them supersymmetric. This is $\mathcal{N}=2$ global SUSY, since we are working in Euclidian signature. 

We further note that these transformations are non-linear; unless the superpotential is quadratic. If it isn't, we can render them linear by introducing auxiliary fields, $F_I=\partial W/\partial\phi_I,$ in terms of which the transformations become linear in the fields, whatever the superpotential, $W:$
\begin{equation}
\label{SUSYtransfsF}
\begin{array}{lcl}
\displaystyle
\delta_{\zeta_1}F_I = \zeta_1\dot{\chi}_I & & 
\displaystyle
\delta_{\zeta_2}F_I = \zeta_2\dot{\psi}_I
\end{array}
\end{equation} 

We remark that the expression~(\ref{Langevin})  for the Nicolai map can be read off the transformation of the anticommuting fields, $\delta_{\zeta_1}\psi_I$ and $\delta_{\zeta_2}\chi_I.$ Therefore, given the supersymmetric theory, we can deduce the Nicolai map from these transformations--at least for the Wess--Zumino class of models. The non-trivial statement here is that there exist two ``noise fields'', non-linear and non-local (in the sense that they contain derivatives) expressions of the scalars, that are free fields, with ultra--local 2-point function. The effort reviewed in ref.~\cite{Lechtenfeld:2023gcq} focuses on the attempt to ``inverse'' this relation and reconstruct the scalars, from the knowledge of the noise fields. This was, already, attempted in refs.~\cite{DeAlfaro:1986uv} and previous papers in the series, without, however,  a conclusive result. 

What the Nicolai map highlights is that, when it can be defined, it can be used to check that the description is {\em complete}, or not:  If the noise fields, defined through the Nicolai map, in terms of the dynamical fields of the theory, are, indeed, free fields, with ultra--local 2-point function, when the correlation functions of the dynamical fields satisfy the Schwinger--Dyson identities, then the description is  complete--and supersymmetry is intact. If the noise fields have non-vanishing 1-point function, but all other identities of the the free fields are satisfied,  then supersymmetry is spontaneously broken. If the identities have anomalies, then supersymmetry is anomalously broken. 

In conclusion, for a non-relativistic particle it can always be defined and no anomalies appear, due to tunneling~\cite{Nicolis:2016osp,Nicolis:2014yka}. Anomalies can appear in the zero-dimensional case~\cite{nicolis_zerkak}, where tunneling is suppressed; this may provide the explanation of the issues raised in ref.~\cite{Golterman:1982pt}.

What happens for the relativistic particle--along the lines of the work by Brink et al.~\cite{Brink:1976uf}--remains to be studied.

Let us now review what changes for field theories. The typical example is the $\mathcal{N}=2,d=2$ Wess--Zumino model, studied by Parisi and Sourlas. There anomalies are not expected to appear.

\subsection{The Nicolai map for field theories}\label{stochSUSY}
The idea, once more, is to postulate the Nicolai map and then show that it is, indeed, as expected. It is here that we encounter something unexpected. 

In two Euclidian spacetime dimensions, we would like to write the Nicolai map as
\begin{equation}
\label{Nicolai2D}
\begin{array}{l}
\displaystyle
\eta_I=a_{IJ}^x\partial_x\phi_J + a_{IJ}^y\partial_y\phi_J+\frac{\partial W}{\partial\phi_I}
\end{array}
\end{equation}
where $I,J=1,2$ and  the 2$\times$2 matrices $a^x$ and $a^y$ are, for the moment, arbitrary numerical coefficients. What Parisi and Sourlas noted was that, {\em if} these are chosen to be   the two Pauli matrices, $a^x=\sigma^x$ and $a^y=\sigma^z,$ {\em then} the 
Jacobian of the change of variables from the noise fields, $\eta_I$ to the scalars $\phi_I,$ 
can be included in the action by anticommuting fields, that have the properties of target space spinors. 

This is a quite remarkable result, whose significance hasn't been fully analyzed. The indices 
$I,J=1,2$ that appear in eq.~(\ref{Nicolai2D}) seem to be ``flavor'' rather than ``spinor'' indices. And the matrices $a^x$ and $a^y$ are numerical coefficients. Nevertheless, they can be used to define target space spinors. 

Furthermore, pursuing the idea of Parisi and Sourlas, we can ask the question, whether we need to include the determinant--therefore the fermionic degrees of freedom--at all, or can it emerge as the expression of the fluctuations, as was the original idea of Parisi and Sourlas. 
Numerical simulations~\cite{Nicolis:2017lqk}  seem to suggest that the answer is affirmative: We do not need to include the fermions at all, their presence emerges as the expression of the fluctuations, as was, also, imagined by Nicolai. 

But we can have a situation in which fermionic degrees of freedom must be assigned a presence, that isn't the expression of fluctuations--what then? The answer is that their flucutuations will be described by superpartners, that are commuting fields--that are correspondingly easier to take into account using numerical simulations, than are fermionic fields.  Therefore there is a ``web of supersymmetries'' that describes such systems. The details will be presented in forthcoming work.
\section{Conclusions and outlook}\label{concl}
In this contribution we have presented the thesis that it is possible to view supersymmetry in two complementary ways:
\begin{itemize}
\item Supersymmetry is at our discretion; it is a symmetry that is consistent with the known symmetries that describe natural phenomena, but it isn't an inevitable property of natural phenomena. 
\item Supersymmetry, in fact, does describe an inevitable property of natural phenomena, in the case of physical systems, in equilibrium with a bath of fluctuations. In that case it provides the framework for describing the degrees of freedom that can resolve the bath of fluctuations. 
\end{itemize}
While much of the effort in understanding supersymmetry has focused on the first option, the thesis of this contribution is that the second option deserves to be taken seriously and passes, already, many non-trivial tests~\cite{Nicolis:2016osp,Nicolis:2014yka,Nicolis:2017lqk}. The relation between the two, indeed, is provided by the Nicolai map, whose scope is much broader than hitherto acknowledged. While for Wess-Zumino models it is now understood how to construct it and how to describe the anomalies it highlights~\cite{Nicolis:2021buh,Nicolis:2023mre,Nicolis:2024qrn}, much remains to be understood, in particular, regarding gauge theories-revisiting the work by Elitzur, Rabinovici and Schwimmer~\cite{Elitzur:1982vh} seems promising, especially in light of the way the stochastic identities allow us unexplored leeway for describing anomalies, going beyond what has been hitherto explored~\cite{Kuzenko:2019vvi}.

However, how to leverage this knowledge of extended supersymmetry to describe theories with chiral fermions remains a challenge. There are conceptual issues that need to be resolved. This is where insights from partial supersymmetry breaking~\cite{Kiritsis:1997ca}
may prove useful.

{\bf Acknowledgements:} I'd like to thank the organizers of the Corfu Workshop for a wonderful conference and M. Axenides, C. P. Bachas, P. Fayet, E. G. Floratos, J. Iliopoulos and A. Schwimmer  for most informative discussions.

\bibliographystyle{/Users/nicolis/Documents/AdS2fromAdS2N/JHEP}
\bibliography{/Users/nicolis/Documents/SUSY/SUSY}

\end{document}